\documentclass[%
reprint,
showpacs,
bibnotes,
amsmath,
amssymb,
aps,
pra,
floatfix
]{revtex4-1}

\usepackage{graphicx}% Include figure files
\usepackage{dcolumn}% Align table columns on decimal point
\usepackage{bm}% bold math
\usepackage{hyperref}
\usepackage{multirow}
\usepackage{array}
\usepackage{booktabs}
\usepackage{ctable}
\usepackage{upgreek}
\usepackage{epsfig}
\usepackage{mathrsfs}
\usepackage{amssymb}
\usepackage{amsbsy}
\usepackage{color}
\usepackage{cancel}
\usepackage{marginnote}
\newcommand{\bcen}{\begin{center}}
\newcommand{\ecen}{\end{center}}
\newcommand{\btab}{\begin{tabular}}
\newcommand{\etab}{\end{tabular}}
\newcommand{\bdes}{\begin{description}}
\newcommand{\edes}{\end{description}}

\newcommand{\beq}{\begin{equation}}
\newcommand{\eeq}{\end{equation}}
\newcommand{\bea}{\begin{eqnarray}}
\newcommand{\eea}{\end{eqnarray}}
\newcommand{\non}{\nonumber}

\newcommand{\half}{\frac{1}{2}}
\newcommand{\bary}{\begin{array}}
\newcommand{\eary}{\end{array}}
\newcommand{\benum}{\begin{enumerate}}
\newcommand{\eenum}{\end{enumerate}}
\newcommand{\bitem}{\begin{itemize}}
\newcommand{\eitem}{\end{itemize}}

\newcommand{\btau}{\mbox{\boldmath $ \tau $}}
\newcommand{\blam}{{\boldsymbol{\lambda}}}

\newcommand{\bPsi}{{\boldsymbol \Psi}}
\newcommand{\bOne}{{\boldsymbol 1}}
\newcommand{\be} { \mbox{\boldmath $e$}}

\newcommand{\bk} { \bm{k} }

\newcommand{\bp} { \bm{p} }

\newcommand{\br} { \boldsymbol{r}}

\newcommand{\bv} { \mbox{\boldmath $v$}}

\newcommand{\bA} { \mbox{\boldmath $A$}}

\newcommand{\dou}{\partial}

\newcommand{\D}[1]{\mbox{d}{#1}}

\newcommand{\mean}[1]{\langle #1 \rangle}
\newcommand{\bra}[1]{{\langle #1 |}}
\newcommand{\ket}[1]{| #1 \rangle}
\newcommand{\braket}[2]{\langle #1 | #2 \rangle}

\newcommand{\eqn}[1] {eqn.~(\ref{#1})}
\newcommand{\prn}[1] {(\ref{#1})}

\newcommand{\Fig}[1]{Fig.~\ref{#1}}

\makeatletter

\newcommand{\Rmnum}[1]{\expandafter\@slowromancap\romannumeral #1@}
\makeatother

\newcommand{\myfigwidth}{0.4\paperwidth}

\newcommand{\Ef}{E_F}
\newcommand{\kf}{k_F}

\newcommand{\bhatp}{{\hat{\bp}}}
\newcommand{\hatp}{{\hat{p}}}
\newcommand{\alphaofp}{{\ket{\chi_\alpha(\bp)}}}
\newcommand{\plusofp}{{\ket{\chi_+(\bp)}}}
\newcommand{\omegazero}{{\omega_0}}
\newcommand{\Heff}{{{\cal H}_{\mbox{{\scriptsize eff}}}}}

\newcommand{\myonlinecite}[1]{[\onlinecite{#1}]}

\newcommand{\mylabel}[1]{\label{#1}} 

\usepackage{lineno}
\begin{document}

%\preprint{}

%\preprint{}
\title{Trapped  fermions in a synthetic non-Abelian gauge field}
%\date{August 24, 2011}
\author{Sudeep Kumar Ghosh}
\email{sudeep@physics.iisc.ernet.in}
\author{Jayantha P. Vyasanakere}
\email{jayantha@physics.iisc.ernet.in}
\author{Vijay B. Shenoy}
\email{shenoy@physics.iisc.ernet.in}
% %Lines break automatically or can be forced with \\
%\thanks{VBS, {\tt shenoy@physics.iisc.ernet.in}}%
%\email{shenoy@physics.iisc.ernet.in}
\affiliation{Centre for Condensed Matter Theory, Department of Physics, Indian Institute of Science, Bangalore 560 012, India}


%\date{\today}

\begin{abstract}

On increasing the coupling strength ($\lambda$) of a non-Abelian gauge field
that induces a generalized Rashba spin-orbit interaction, the topology
of the Fermi surface of a homogeneous gas of noninteracting fermions
of density $\rho \sim \kf^3$ undergoes a change at a critical value, $\lambda_T \approx \kf$ [Phys.~Rev.~B
  {\bf 84}, 014512 (2011)]. In
this paper we analyze how this phenomenon affects the size and shape
of a cloud of spin-$\half$ fermions trapped in a harmonic potential such
as those used in cold atom experiments. We develop an adiabatic
formulation, including the concomitant Pancharatnam-Berry phase
effects, for the one particle states in the presence of a trapping
potential and the gauge field, obtaining approximate analytical
formulae for the energy levels for some high symmetry gauge field
configurations of interest. An analysis based on the local density
approximation reveals that, for a given number of particles, the cloud shrinks in a {\em characteristic fashion with
increasing $\lambda$}. We explain the physical origins of this effect
by a study of the stress tensor of the system. For an isotropic
harmonic trap, the local density approximation predicts a spherical
cloud for all gauge field configurations, which are anisotropic in general. We show, via a calculation of
the cloud shape using exact eigenstates, that for certain gauge field
configurations there is systematic and observable anisotropy in the
cloud shape that increases with increasing gauge coupling
$\lambda$. The reasons for this anisotropy are explained using the
analytical energy levels obtained via the adiabatic
approximation. These results should be useful in the design of cold
atom experiments with fermions in non-Abelian gauge fields. An
important spin-off of our adiabatic formulation is that it reveals
exciting possibilities for the cold-atom realization of interesting
condensed matter Hamiltonians by using a non-Abelian gauge field in
conjunction with another potential. In particular, we show that use of
a spherical non-Abelian gauge field with a harmonic trapping
potential produces a spherical geometry quantum hall like Hamiltonian
in the momentum representation.

\end{abstract}

\pacs{03.75.Ss, 05.30.Fk, 67.85.Lm}

\maketitle

\section{Introduction}
\mylabel{sec:Introduction}

Developments in the area of cold atoms in the past
decade\cite{Ketterle2008,Bloch2008,Giorgini2008} have opened the possibility of
obtaining insights into many outstanding problems of condensed matter
physics using cold atom quantum emulators. While this program has
met with early success, serious difficulties have hampered rapid experimental
progress. Notable roadblocks are the problem of entropy removal
(``cooling'')  and the
generation of magnetic fields necessary for the attainment of
interesting states such as the correlated quantum hall states.

There have been many theoretical
proposals\cite{Jaksch2003,Osterloh2005,Ruseckas2005,Gerbier2010} for the generation of both
Abelian and non-Abelian gauge fields. Interest in this problem has
resurged owing to recent experimental progress from the NIST
group in the generation of synthetic gauge
fields (both Abelian and non-Abelian) using bosonic $^{87}$Rb atoms.\cite{Lin2009A, Lin2009B, Lin2011} 
A spatially inhomogeneous Abelian gauge field produces non-trivial
effects of a magnetic field, while a uniform Abelian gauge field is
tantamount to a mere gauge transformation. On the other hand even a
uniform non-Abelian gauge field produces interesting physical
effects. In the context of bosons some aspects of these have been
investigated and reported.\cite{Ho2010,Wang2010,Lin2011}

A natural question that arises pertains to the effect of a uniform
non-Abelian gauge field on fermions. Such a gauge field induces a generalized
Rashba spin-orbit interaction. The hint that such a system contains
interesting physics came from ref.~\myonlinecite{Vyasanakere2011}
which demonstrated that a high symmetry (see below)
non-Abelian gauge field induces a bound state between two fermions for
{\em any} attractive interaction however small (small negative
scattering length).  This motivated the study\cite{Vyasanakere2011b}
of the evolution of the many body fermionic ground state with the
increasing strength $\lambda$ of the
non-Abelian gauge field. The ground state realized for a weak
attractive interaction in the absence of the gauge field is the BCS
(Bardeen-Cooper-Schrieffer) superfluid state with large overlapping
Cooper pairs. Ref.~\myonlinecite{Vyasanakere2011b} demonstrated that
on increasing the gauge coupling strength $\lambda$ the ground state evolves
from this just discussed BCS state to a {\em BEC (Bose-Einstein
  condensate) state}, i.~e., the non-Abelian gauge field induces a
BCS-BEC crossover even at a fixed weak attraction that is unable to
produce a bound state between two fermions in
the absence of the gauge field. The most remarkable aspect of the
crossover induced by the gauge field is that the BEC state obtained
for large gauge couplings is made up of a new kind of tightly bound
fermion-pairs called\cite{Vyasanakere2011b} 
``rashbons''. Hence, the condensate is called as rashbon-BEC (RBEC). Rashbons are
anisotropic bosons characterized by an anisotropic dispersion and spin
structure (nematicity), all of which are determined {\em solely} by the
gauge field. The properties of RBEC, in particular the transition
temperature, is determined by those of the corresponding
rashbons. Properties of rashbons are reported in
ref.~~\myonlinecite{Vyasanakere2011c}, where it is demonstrated that a
non-Abelian gauge field can enhance the exponentially small transition
temperature of a Fermi gas with a weak attraction to the order of the Fermi temperature. Aspects of rashbon
superfluidity, effects of Zeeman field and effects of imbalance etc. have been reported in
refs.~\cite{Gong2011,Hu2011,Yu2011,Iskin2011A}.

A non-Abelian gauge field also has quite interesting effects on {\em
  noninteracting} fermions. In fact, for a fixed density $\rho$ of
the particles, increasing the gauge coupling strength $\lambda$
induces a change in the topology of the Fermi surface at a critical gauge coupling strength
$\lambda_T$. As shown in ref.~\myonlinecite{Vyasanakere2011b},
$\lambda_T$ is always of the order of $\kf$, the Fermi wave vector
determined by the density ($\rho = \kf^3/(3 \pi^2)$). In fact, in the
presence of a weak attraction, the regime of gauge coupling over which
the crossover from the BCS to RBEC takes place coincides with $\lambda
\gtrsim \lambda_T$. In our opinion, one of the first things cold atom
experiments designed to study fermions in non-Abelian gauge fields
should probe is this transition in the topology of the Fermi surface.
The discussion above naturally gives rise to many interesting
questions. Since almost all cold atom experiments are performed in a
trap, it is useful to know the key signatures that provide
measurable/falsifiable proof of the physics discussed above. A
specific question is: How the presence of a generic non-Abelian gauge
field influence the size and shape of a cloud of fermions? Answering
this question is the key motivation of this paper.

In this paper we investigate noninteracting fermions in a parabolic trapping potential characterized by a scale $\omegazero$, in presence of a synthetic non-Abelian gauge field (whose strength is characterized by $\lambda$). Our goal is to understand how a synthetic
non-Abelian gauge field affects the size and shape of a cloud of
noninteracting fermions. To this end: (a) Noting that the most
interesting regime corresponds to $\omegazero \ll \lambda^2$, we develop
a Born-Oppenheimer like approximation for the states of a
trapped particle. This analysis reveals how the internal fast degree
of freedom induces a Berry connection, i.~e., a gauge
potential on the motion of the particle, and most importantly {\em
  suggests routes to generating interesting states} such as the
spherical geometry quantum hall states using cold atoms. (b) We study
the size and shape of the cloud of trapped noninteracting fermions
under the influence of a non-Abelian gauge field using the local
density approximation (LDA). (c) Finally, we compare the results of the
LDA with exact numerical calculation. This last study reveals {\em systematic} and observable deviations from the LDA which are explained using the result of our adiabatic theory.
This study finds that the cloud shrinks (consistent with ref.~\cite{Hu2011}). Importantly we uncover the scaling of the cloud size and the evolution of the anisotropy of the density profile with increasing gauge coupling, characteristic to various gauge field configurations of interest. We believe this will be of value for experiments on fermions in a non-Abelian gauge field.
 An important byproduct of our adiabatic analysis is the possibility of realization of interesting physics in cold atomic systems such as the spherical geometry quantum hall state by using a synthetic non-Abelian gauge field in conjunction with another potential.

Section \ref{sec:Preliminaries} consists of the background including notation, estimation of scales and sets up the statement of the problem. This is followed by a discussion of the one particle states in section \ref{sec:OnePStates}. The effect of the gauge field on the cloud size and shape is discussed in section~\ref{sec:CloudSizesAndShapes}. The final section \ref{sec:Summary} consists of an itemized summary of the paper. Appendix \ref{sec:NumericalCalculation} outlines the numerical method used to obtain exact one particle states for an extreme oblate gauge field configuration.

\section{Preamble and problem statement}
\mylabel{sec:Preliminaries}

Denoting $\bPsi^\dagger(\br) = \{\Psi_{\sigma}(\br)\}, \sigma= \uparrow, \downarrow$  as creation operators of spin-$\half$ fermions at position $\br$ in three spatial dimensions, the Hamiltonian under consideration (see refs.~\cite{Vyasanakere2011, Vyasanakere2011b} for more details) is
\beq\mylabel{eqn:Rashba}
{\cal H}_R = \int \D{ \br} \, \bPsi^\dagger(\br) \left( \frac{\bhatp^2}{2} \bOne - \bhatp_\lambda \cdot \btau \right) \bPsi(\br),
\eeq
where $\bhatp$ is the momentum operator, $\bOne$ is the SU(2) identity, $\tau^\mu$ ($\mu=x,y,z$) are Pauli matrices, $\bhatp_\lambda = \sum_i \hatp_i \lambda_i \be_i$, $\be_i$'s are the unit vectors in the $i$-th direction, $i=x,y,z$. The gauge-field configuration (GFC) is described by a vector $\blam =\sum_i \lambda_i \be_i$, and $\lambda = |\blam|$ is the gauge-coupling strength. We work in units where the Planck constant ($\hbar$) and the fermion mass $m$ are set to unity. We are particularly interested in high symmetry GFCs called extreme oblate (EO) GFC with $\blam = \frac{\lambda}{\sqrt{2}} ( \be_x + \be_y)$ and spherical (S) GFC with $\blam = \frac{\lambda}{\sqrt{3}} (\be_x + \be_y + \be_z)$, which nurture interesting physics.\cite{Vyasanakere2011,Vyasanakere2011b}
 
The eigenstates of the noninteracting Hamiltonian (\eqn{eqn:Rashba}) are
\beq\mylabel{eqn:HelicityEigenstates}
\ket{\bp\alpha} = \ket{\bp} \otimes \alphaofp
\eeq
where $\ket{\bp}$ is the plane wave state with momentum eigenvalue $\bp$, $\alpha=\pm 1$ is the helicity with 
\beq\mylabel{eqn:HelicityOperator}
\bhatp_\lambda \cdot \btau \alphaofp = \alpha |\bp_\lambda| \alphaofp.
\eeq
In other words, $\alphaofp$ is the eigenstate of $\bhatp_\lambda \cdot \btau$ in the spin sector with associated helicity $\alpha$. The energy eigenvalues associated with the states in \eqn{eqn:HelicityEigenstates} are
\beq\mylabel{eqn:EnergyEigenvalue}
\varepsilon_{\alpha}(\bp) = \frac{p^2}{2} - \alpha |\bp_\lambda|.
\eeq

We now introduce an isotropic harmonic trapping potential
\beq\mylabel{eqn:TrappingPotential}
{\cal H}_T = \frac{\omegazero^2}{2} \int \, \D{\br} \, r^2  \bPsi^\dagger(\br) \bPsi(\br),
\eeq
where $\omegazero$ is the trapping frequency. Unfortunately, the trapping potential spoils the symmetries of the system. Not only does it not commute with the usual kinetic energy term, it also does not let the helicity $\alpha$ of \eqn{eqn:HelicityOperator} to be a good quantum number. This renders the diagonalization the full Hamiltonian  
\beq\mylabel{eqn:FullHamiltonian}
{\cal H} = {\cal H}_R + {\cal H}_T
\eeq
for a generic $\omegazero$ and $\blam$ analytically intractable.

Fortunately, there are two well separated energy scales in the problem in the regime of interest. To see this, the typical traps used in experiments are such that the Fermi energy (set by the density of particles  at the trap center) denoted as $\Ef^0 = {\kf^0} ^2/2$ and the trap frequency satisfies
\beq
\frac{\Ef^0}{\omegazero} \approx 10^2.
\eeq
We have also noted that the regime of interest of the gauge coupling corresponds to $\lambda \gtrsim \lambda_T$ where a change in the topology of the Fermi surface is engendered by the gauge field. In this regime of gauge coupling all the occupied states are of positive helicity. Noting that $\lambda_T \approx \kf^0$ for GFCs of interest, we see that the regime of interest corresponds to 
\beq\mylabel{eqn:InterestingRegime}
\frac{\lambda^2}{\omegazero} \gtrsim 10^2
\eeq
In the next section, we show that the regime indicated in \eqn{eqn:InterestingRegime} allows us to make the adiabatic approximation and obtain the energy levels. 

\section{One Particle States -- Adiabatic theory}
\mylabel{sec:OnePStates}

\subsection{Adiabatic Approximation for a Generic GFC}
\mylabel{sec:Adiabatic}
The development of the adiabatic approximation to the diagonalization of \eqn{eqn:FullHamiltonian} begins with the classification of the slow and fast degrees of freedom. This is most conveniently done in first quantized notation in {\em momentum} representation:
\beq\mylabel{eqn:FirstQuantuntized}
{\cal H} = - \frac{\omegazero^2}{2} \frac{\dou^2 }{\dou \bp^2} + \frac{p^2}{2} - \bp_\lambda \cdot \btau
\eeq
where we have used $\hat{\br} = i \frac{\dou}{\dou \bp}$, the position operator in momentum representation.  
In the regime of interest (\eqn{eqn:InterestingRegime}), $\lambda \gg \sqrt{\omegazero}$, the internal spin degree of freedom is the ``fast variable'', while the orbital degree of freedom is the ``slow variable''. This identification of the fast and slow degrees of freedom suggests an ansatz for the wave function of the particle
\beq\mylabel{eqn:BOansatz}
\ket{\psi} = \int \, \D{ \bp} \, \psi(\bp) \, \ket{\bp} \otimes \alphaofp.
\eeq
The main physical content of this approximation is that the particle motion is such that with the change of $\bp$, it instantly attains internal spin state corresponding to a given helicity $\alpha$ associated with $\bp$, hence its internal state is $\alphaofp$. We shall restrict attention to $\alpha=+1$ since this is the relevant case for $\lambda \gtrsim \lambda_T$.

The goal now is to find the effective Hamiltonian for the wave function $\psi(\bp)$. Since the spin Hilbert space has nontrivial topology, the associated Pancharatnam-Berry phase\cite{Berry1989} induces a connection which results in a new gauge field and a potential for the slow degree of freedom $\bp$. Following the line of analysis of ref.\cite{Berry1989}, we obtain the effective Hamiltonian as
\beq\mylabel{eqn:EffectiveHamiltonian}
\Heff  = \frac{\omegazero^2}{2} \left(i \frac{\dou}{\dou \bp} - \bA \right)^2 + \varepsilon_\alpha(\bp) + V_{BO}(\bp)
\eeq
where
\beq\mylabel{eqn:GaugeField}
\bA = - i \braket{\chi_\alpha(\bp)}{\frac{\dou \chi_\alpha(\bp)}{\dou \bp}}
\eeq
is the induced connection (U(1) gauge potential) and
\beq\mylabel{eqn:BOPotential}
\begin{split}
V_{BO}(\bp) & = \frac{\omegazero^2}{2} \left(\braket{\frac{\dou \chi_\alpha(\bp)}{\dou p_i}}{\frac{\dou \chi_\alpha(\bp)}{\dou p_i}} \right. \\
& \left.  - \braket{\frac{\dou \chi_\alpha(\bp)}{\dou p_i}}{\chi_\alpha(\bp)}  \braket{\chi_\alpha(\bp)}{\frac{\dou \chi_\alpha(\bp)}{\dou p_i}}  \right)
\end{split}
\eeq
is a potential where repeated spatial indices are summed. The energy levels are now obtained from the eigenvalue equation
\beq\mylabel{eqn:EffectiveEigenvalue}
\Heff \psi(\bp) = \varepsilon \psi(\bp).
\eeq
It is quite interesting to note that adiabatic motion in a non-Abelian gauge field results in a U(1) (Abelian) gauge field for the slow degree of freedom. Clearly, this promises to give rise to new possibilities with cold atom systems. We shall come back to this point in the sections that follow.

We now discuss application of the formulae developed in this section to various GFCs of interest.

\subsection{Extreme oblate  (EO) GFC}
\mylabel{sec:EOOneP}

To solve for the levels of \eqn{eqn:EffectiveEigenvalue} for the EO GFC, we choose cylindrical polar coordinates in the momentum space $(p,\phi,p_z)$ with associated unit vectors $\be_{p}$, $\be_\phi$ and $\be_z$. Here we consider the positive helicity state ($\alpha = +1$) and 
\beq
\plusofp \equiv  \frac{1}{\sqrt{2}} 
\left( \begin{array}{c}
1 \\
e^{i \phi}
\end{array}
 \right)
\eeq
We immediately obtain from \eqn{eqn:GaugeField} and \eqn{eqn:BOPotential} that
\beq\mylabel{eqn:EOgaugefield}
\bA = \frac{1}{2 p} \be_\phi, \;\;\;\; V_{BO}(\bp) = \frac{\omegazero^2}{8 p^2}.
\eeq
The gauge field corresponds to a half flux-quantum magnetic field line running along the $z$ axis of the momentum space.
The effective Hamiltonian from \eqn{eqn:EffectiveHamiltonian} reduces to
\beq\mylabel{eqn:EOHeff}
\begin{split}
\Heff & = -\frac{\omegazero^2}{2} \left( \frac{\dou^2}{\dou p^2} + \frac{1}{p} \frac{\dou}{\dou p} \right) + \frac{\omegazero^2}{2 p^2} (L_z + \half)^2 \\ 
      & + \frac{p^2}{2} - \frac{\lambda}{\sqrt{2}} p + \frac{\omegazero^2}{8p^2} \\
      &  -\frac{\omegazero^2}{2} \frac{\dou^2}{\dou p^2_z} +  \frac{p_z^2}{2}
\end{split}
\eeq
where $L_z = -i \frac{\dou}{\dou \phi}$ is the $z$ component of the orbital angular momentum operator.

 We use the separation of variable ansatz $\psi(p,\phi,p_z) = \psi_p(p) \psi_\phi(\phi) \psi_z(z)$, and note that $L_z$ eigenvalue can be labeled by an integer $m$ to obtain 
\beq\mylabel{eqn:WKBeqn}
-\frac{\omegazero^2}{2} u''(p) + \left[\frac{\omegazero^2}{2 p^2} \left(m + \half \right)^2 + \frac{p^2}{2} - \frac{\lambda}{\sqrt{2}} p \right] u(p) = \tilde{\varepsilon} u(p)
\eeq
where $\psi_p(p) = \frac{u(p)}{\sqrt{p}}$ (a standard substitution), and $\tilde{\varepsilon}$ is the ``energy in the $(p,\phi)$ degrees of freedom''. This form is tailor-made for a semi-classical WKB treatment\cite{Brack1997}, which  gives
\beq
\tilde{\varepsilon}(n,m) \approx -\frac{\lambda^2}{4} + \left(n + \half \right) \omegazero + \frac{\omegazero^2}{\lambda^2} \left(m + \half \right)^2
\eeq
where $n$ is the ``radial quantum number''.
Remarkably, the presence of the potential induced by the connection obviates the need for the usual Langer modification \cite{Langer1937,Brack1997} necessary in semi-classical analysis of this type and has the Maslov index of unity at both the turning points. Adding in the part of the energy from the $z$ degree of freedom we obtain the energy eigenvalues of \eqn{eqn:EOHeff} dependent on three quantum numbers to be
\beq\mylabel{eqn:EOEigval}
\varepsilon(n,m,n_z) \approx -\frac{\lambda^2}{4} + (n + n_z + 1) \omegazero +  \frac{\omegazero^2}{\lambda^2} \left(m + \half \right)^2 
\eeq

\begin{figure*}
\centerline{\includegraphics[width=\myfigwidth]{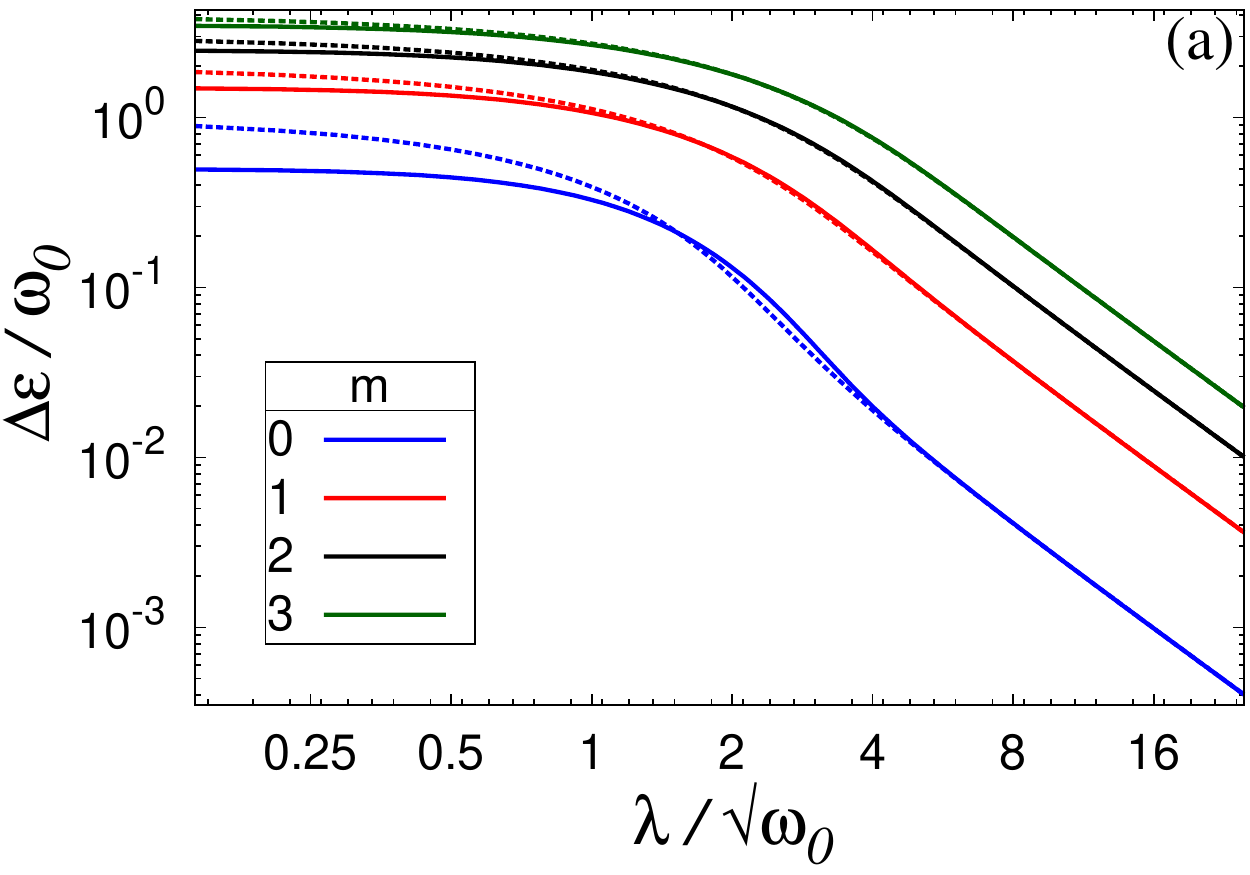}~~~~~\includegraphics[width=\myfigwidth]{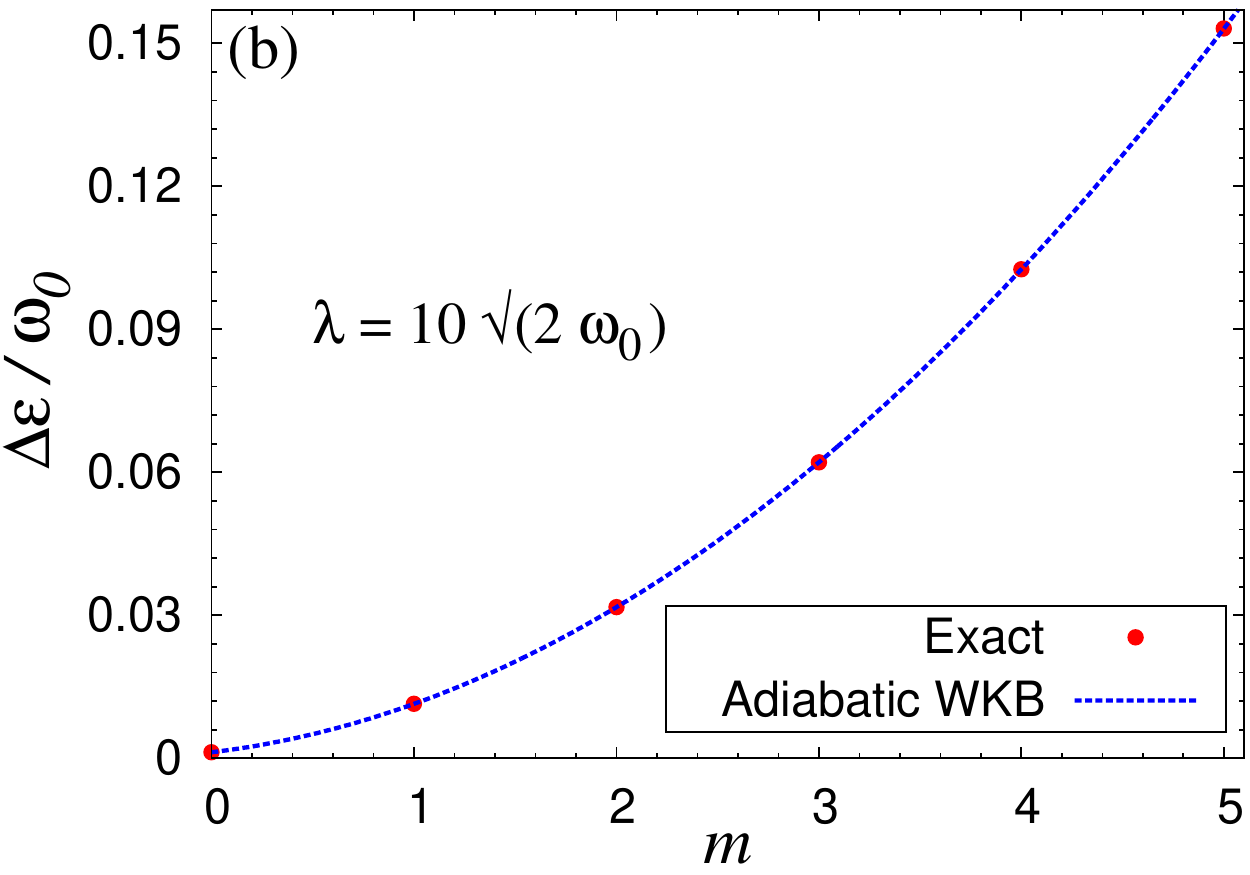}}
\caption{(Color online) One particle levels of a trapped particle in an EO GFC. Comparison of numerically calculated adiabatic WKB energies (dashed lines) obtained from the analysis of \eqn{eqn:WKBeqn} and  the exact result (solid lines in (a) and points in (b))  obtained using the method outlined in appendix \ref{sec:NumericalCalculation}. Here $\Delta \varepsilon = \varepsilon(0,m,0) + \frac{\lambda^2}{4} - \omegazero$. For $\lambda \gtrsim 5 \sqrt{\omegazero}$ the results are indistinguishable with the approximate result of \eqn{eqn:EOEigval}. (a) Energy  for various values of $m$ quantum number as a function of $\lambda$. (b) Energy for $\lambda=10\sqrt{2 \omegazero}$ as a function of the quantum number $m$.}
\mylabel{fig:EOcomparison}
\end{figure*}

We now present a comparison of the result of \eqn{eqn:EOEigval} with the
exact numerical solution of \eqn{eqn:FullHamiltonian} briefly outlined
in appendix~\ref{sec:NumericalCalculation}. \Fig{fig:EOcomparison}(a) shows a plot of
$\Delta\varepsilon = \varepsilon(0,m,0) + \frac{\lambda^2}{4} - \omegazero$ obtained
numerically (solid lines) and the same quantity calculated using full numerical WKB solution (dashed lines) of \eqn{eqn:WKBeqn}. This numerical WKB solution very closely matches the approximate analytical formula (\eqn{eqn:EOEigval}) for  $\lambda > 6\sqrt{\omega_0}$. We see that the full numerical WKB solution has a quite remarkable agreement with the exact solution of \eqn{eqn:FullHamiltonian} even for as small value of gauge coupling as $\lambda = 4 \sqrt{\omegazero}$ for $m=0$, and agreement to even lower values of $\lambda$ for higher values of $m$.

Why does the adiabatic approximation work so well,
and why does it work better for larger $m$? This can be understood by
noting that key ingredient of the adiabatic approximation is to
``force'' the particle to have a fixed helicity. This approximation is
quite valid when the momentum of the particle is large ($p \gtrsim
\lambda$). In the presence of the trapping potential, helicity
fluctuations are maximum for momenta near the origin. Now note that
the gauge potential induced by the Berry connection
effectively provides for a nonzero centrifugal barrier for all $m$
including $m=0$ in momentum space (see \eqn{eqn:WKBeqn}, term
proportional to $(m+ \half)^2$). This effectively ``keeps the
particle away from the origin'' in momentum space, hence minimizing
helicity fluctuations and rendering the adiabatic approximation
accurate. The same argument provides the reason why the approximation
works better for larger $m$. In the discussion that follows we shall
use \eqn{eqn:EOEigval}. We note that a similar formula was arrived at
by different considerations in refs.~\myonlinecite{Hu2011b,Sinha2011}
that appeared during the preparation of this manuscript.

\subsection{Spherical (S) GFC}
With the reassuring success of the adiabatic approximation for the EO GFC, we now turn to spherical GFC. The natural coordinate system to treat this GFC via \eqn{eqn:EffectiveHamiltonian} is the spherical polar coordinate system in momentum space with $(p,\theta,\phi)$ as the coordinates and $\be_p,\be_\theta$ and $\be_\phi$ as the basis vectors. The positive helicity eigenstates have
\beq
\plusofp \equiv \left( \begin{array}{c} 
\cos{\frac{\theta}{2}} e^{- i \phi/2} \\
\sin{\frac{\theta}{2}} e^{ i \phi/2} 
\end{array}
 \right).
\eeq
It follows from \eqn{eqn:GaugeField} and \eqn{eqn:BOPotential} that
\beq\mylabel{eqn:SGFandVBO}
\bA = - \frac{\cot{\theta}}{2 p} \be_\phi,\;\;\;\; V_{BO}(\bp) = \frac{\omegazero^2}{4 p^2}.
\eeq
Quite interestingly, the gauge potential $\bA$ corresponds to {\em the presence of a monopole of charge $Q=\half$ at the origin of the momentum space}. The angular motion (i.~e., $(\theta,\phi)$) of a particle in a monopole field has been extensively studied\cite{Wu1976,Wu1977} and also used in the context of quantum Hall effect in the so called spherical geometry (see section 3.10 of \cite{Jain2007}). We thus obtain,
\beq\mylabel{eqn:SHeff}
\begin{split}
\Heff & = -\frac{\omegazero^2}{2} \left(\frac{1}{p^2} \frac{\dou}{\dou p} p^2 \frac{\dou}{\dou p} \right) \\ 
& + \frac{\omegazero^2}{2 p^2} \left[-\frac{1}{\sin{\theta}} \frac{\dou}{\dou \theta} \sin{\theta} \frac{\dou}{\dou \theta} + \left(Q \cot{\theta} + \frac{i}{\sin{\theta}} \frac{\dou}{\dou \phi} \right)^2 \right] \\
& + \frac{\omegazero^2}{4 p^2} + \left(\frac{p^2}{2} - \frac{\lambda}{\sqrt{3}} p \right).
\end{split}
\eeq
Again using a separation of variable ansatz, $\psi(p,\theta,\phi) = \psi_p(p) \Omega(\theta,\phi)$, and noting that the eigenvalues of the angular operator in the square bracket in \eqn{eqn:SHeff} is $\ell(\ell+1) - Q^2$ with $\ell = |Q|,  |Q|+1, \cdots$ where $\ell$ is the angular momentum quantum number. For each $\ell$ there are $2 \ell +1 $ states with magnetic quantum numbers $m = - \ell, -\ell+1, \dots , \ell$, and the wave function $\Omega(\theta,\phi)$ is one of the monopole harmonics.\cite{Jain2007} We thus obtain
\beq
\begin{split}
& -\frac{\omegazero^2}{2} \left(\frac{1}{p^2} \frac{\dou}{\dou p} p^2 \frac{\dou \psi_p}{\dou p} \right)  \\
& + \left[\frac{\omegazero^2}{2 p^2} \left(\ell( \ell +1) - |Q|^2 + \half \right) +  \left(\frac{p^2}{2} - \frac{\lambda}{\sqrt{3}} p \right)  \right] \psi_p = \varepsilon \psi_p
\end{split}
\eeq
which is amenable to a semi-classical WKB treatment via $\psi_p(p) = u(p)/p$. An approximate analysis gives us
\beq\mylabel{eqn:SWKB}
\varepsilon(n,\ell,m) \approx -\frac{\lambda^2}{6} + (n + \half) \omegazero + \frac{3 \omegazero^2}{2 \lambda^2} \left( \ell (\ell+1) + \frac{1}{4} \right).
\eeq
This result, again, is obtained without the usual Langer modification applied to WKB treatment of three dimensional problems.\cite{Langer1937}

A key inference that can be made from the results of this section is
that non-Abelian gauge fields used in conjunction with a trapping
potential can give rise to many new possibilities with cold atoms. In
particular when the gauge coupling is strong compared to the trapping
potential, the motion of the particle is adiabatic with respect to the
spin degree of freedom. The associated Pancharatnam-Berry phase
produces an effective gauge field which can be used to realize
Hamiltonians that are quite interesting. In this regard, it is very
interesting to note that the spherical non-Abelian gauge field in
conjunction with a harmonic trapping potential just discussed produces
a ``spherical geometry'' quantum hall like Hamiltonian in the momentum
space. Exploitation of this aspect of non-Abelian gauge fields should
open up very interesting possibilities with cold atoms.

\section{Sizes and Shapes of Trapped Clouds}
\mylabel{sec:CloudSizesAndShapes}

In this section we investigate the natural question that arises: Is
there any discernible effect on the size and shape of the trapped
clouds due to the influence of the non-Abelian gauge field? In the following subsection we answer this question using the local density approximation and compare it with the exact calculation in the subsection that follows.
We restrict our attention to zero temperature.

\subsection{Local Density Approximation}
\mylabel{sec:LDA}

In the LDA \cite{Pethick2004}, the spatial dependence of the chemical potential $\mu$ is determined by the trapping potential,
\beq
\mu(\br) + \half \omegazero^2 r^2 = \mu_0 
\label{eqn:LDAmu}
\eeq
where $\mu_0$ is the chemical potential at the trap center, and $\mu(\br)$ is the ``local chemical potential'' that determines the density $\rho(\br)$ of fermions at the point $\br$. It follows from \eqn{eqn:LDAmu} that within LDA, $\mu(\br)$ and hence $\rho(\br)$ will be dependent only on $r=|\br|$. The density $\rho$ is related to the chemical potential $\mu$ by the equation of state. For the GFCs of interest, we can obtain the equation of state analytically. We shall focus mainly on the EO GFC for which the equation of state is
\begin{widetext}
\bea
\rho(\mu) = \begin{cases}
\frac{(2 \mu)^{3/2}}{3 \pi^2} + \frac{\lambda}{2 \sqrt{2} \pi^2} \left(
\lambda \sqrt{\mu} +  \left( \frac{\lambda^2}{2} + 2 \mu \right) \sin^{-1}\left(\frac{\lambda}{\sqrt{4 \mu + \lambda^2}} \right)  \right) & \text{if } \mu \geq 0, \\
\frac{\lambda (4 \mu + \lambda^2)}{8 \sqrt{2} \pi} & \text{if } -\frac{\lambda^2}{4} \leq \mu < 0.
\end{cases} \non
\eea
\end{widetext}

The system is such that the radius of the cloud at $\lambda =0$ is $R_0$, with a trap center density given by $\rho_0= (\kf^0)^3/3 \pi^2$ and an associated Fermi energy $\Ef^0 = (\kf^0)^2/2=\frac{1}{2} \omegazero^2 R_0^2 $. On application of the gauge field, the radius of the cloud changes to $R$. We seek to determine the dependence of $R$ on $\lambda$. This is determined from the equation
\beq
4 \pi \int_0^R \D{r} \; r^2 \rho(\mu(r)) = \frac{1}{24} R_0^6 \omegazero^3
\eeq
where the right hand side is the number of particles in the trap expressed in terms of the radius $R_0$ and the trap frequency $\omegazero$.

\begin{figure}
\centerline{\includegraphics[width=\myfigwidth]{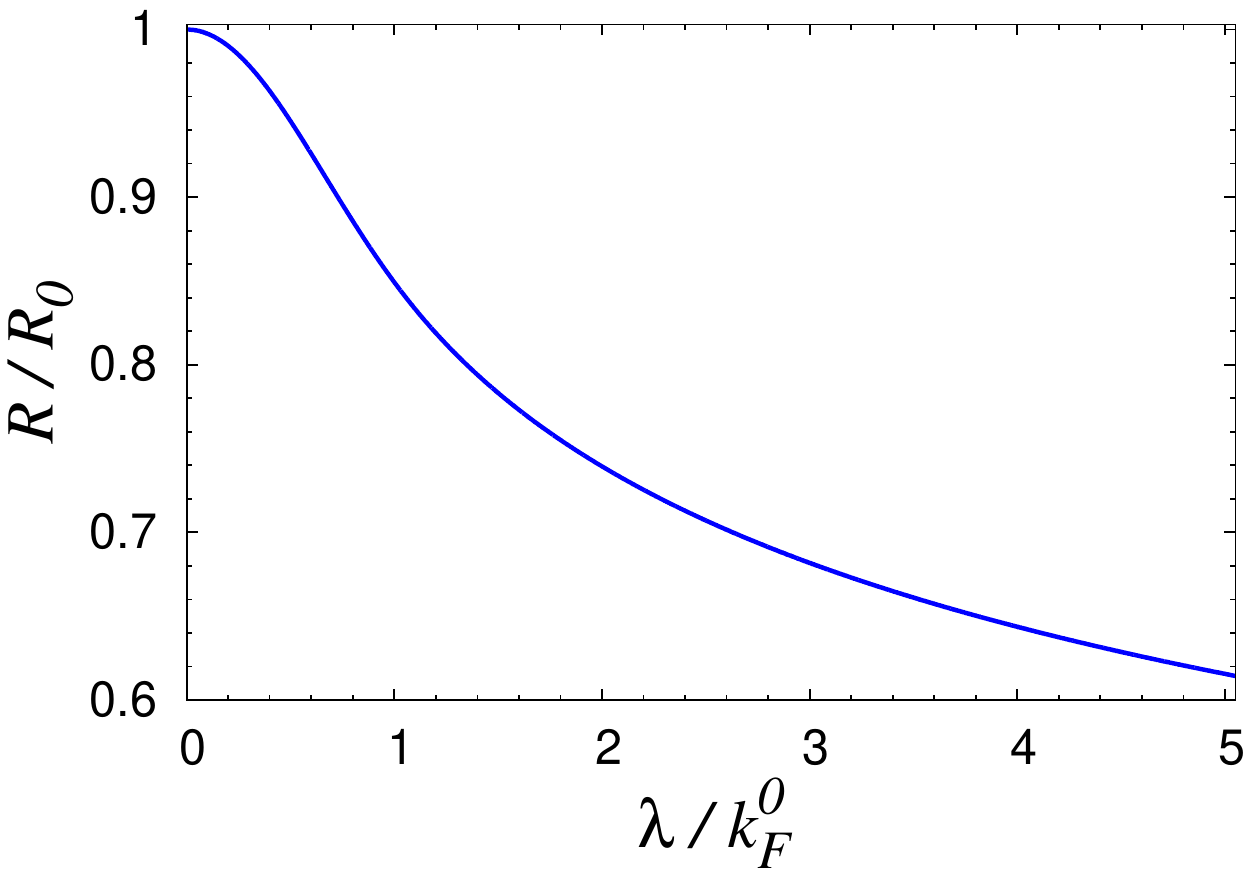}}
\caption{(Color online) Dependence of the radius of the cloud on the gauge coupling $\lambda$ for the EO GFC.  $R_0$ is the radius of the cloud in the absence of the gauge field and  $\kf^0$ is the Fermi wave vector related to the trap center density in the absence of the gauge field. }
\mylabel{fig:EOCloudRadius}
\end{figure}

\Fig{fig:EOCloudRadius} shows a plot of the dimensionless cloud radius $R/R_0$ as a function of the dimensionless gauge coupling strength $\lambda/\kf^0$ of the EO gauge field. Quite remarkably, the cloud shrinks on application of the gauge field. This is consistent with the results of ref.~\cite{Hu2011}. Further, we find here that there is a critical gauge coupling $\lambda_c$ given by
\beq
\lambda_c = \left(\frac{5}{2} \right)^{1/6} \kf^0 ,
\eeq
such that when $\lambda > \lambda_c$, the radius is given by
\beq
\frac{R}{R_0} = \left(\frac{5 \sqrt{2}}{16} \frac{\kf^0}{\lambda} \right)^{1/5}\;\;\;\;\mbox{for}\;\;\; \lambda \ge \lambda_c
\eeq
This is the regime of gauge coupling that has the most interesting physics. This is because for $\lambda > \lambda_c$ the local Fermi sea every where in the trap will contain only $+$ helicity eigen states. In the presence of attractive interaction between fermions in the singlet channel, a rashbon BEC is obtained at the center of the trap at low temperatures when $\lambda \gtrsim \lambda_c$.

For the S GFC, briefly, the cloud radius satisfies 
\beq
\left( \frac{R}{R_0} \right)^6 + 2 \left(\frac{\lambda}{\kf^0}\right)^2 \left( \frac{R}{R_0} \right)^4 = 1.
\eeq
Indeed, even for this GFC, the cloud shrinks under the influence of the gauge field. As is evident, for $\lambda/\kf^0 \gtrsim 1$, the cloud radius goes as $\frac{R}{R_0} \sim \sqrt{\frac{\kf^0}{\lambda}}$.

The discussion above makes it clear that the shrinking of the cloud is
a characteristic feature obtained on application of a generic non-Abelian gauge
field. It is useful to further investigate the physics behind this
remarkable and readily observable effect. Let us briefly recap the
factors that determine the cloud size. In a Fermi system, the
``confining walls'' of the container resist the Pauli (degeneracy)
pressure. In this case equivalent to the confining wall is the
harmonic trapping potential. It is apparent that on application of the
gauge field the degeneracy pressure reduces, an inference that is consistent with
the reduction of the cloud size.

Why should the Pauli pressure of the system reduce in the presence of a
non-Abelian gauge field? To answer this question, we consider a {\em homogeneous system} (no trap) and obtain the stress tensor from the momentum balance equation. For our noninteracting  system, the stress tensor operator is (sum over repeated indices is implied)
\beq\mylabel{eqn:StreeTensorOperator}
S^{i j} = \frac{1}{8 \pi^3} \int \, \D{ \bk} \, \Psi^\dagger_\sigma(\bk) \, k^i v^j_{\sigma \sigma'} (\bk) \, \Psi_{\sigma'}(\bk)
\eeq
where $\Psi^\dagger_\sigma(\bk)$ is the Fermion creation operator in $\bk$ space, 
\beq\mylabel{eqn:Velocity}
v^j_{\sigma \sigma'} (\bk) = \left( k^j \delta_{\sigma \sigma'} - \lambda^{(j)} \tau^{(j)}_{\sigma \sigma'} \right) 
\eeq
are the components of the {\em velocity} operator (here repeated $j$ is not summed). Note that the velocity operator does {\em not} commute with the Hamiltonian in \eqn{eqn:Rashba}. A straightforward calculation now shows that the stress tensor in our case is
\beq\mylabel{eqn:StressTensorExpect}
\mean{S^{ij}} =  \frac{1}{8 \pi^3}  \int \, \D{ \bk} \,  \left( \sum_\alpha k^i \mean{v^j(\bk)}_\alpha n_F(\varepsilon_\alpha(\bk) -\mu) \right)
\eeq
where $n_F$ is the Fermi function, $\mu$ is the chemical potential and $\mean{v^j(\bk)}_\alpha$ is the expectation value of the velocity operator (\eqn{eqn:Velocity}) in the helicity eigenstate $\ket{\bk \alpha}$ given in \eqn{eqn:HelicityEigenstates}. The pressure $P$ is now obtained as
\beq\mylabel{eqn:Pressure}
P = \frac{1}{3} \mean{S^{ii}}.
\eeq

For the EO gauge field,
\beq
\mean{\bv(\bk)}_\alpha = (k - \frac{\alpha \lambda}{\sqrt{2}}) \be_k + k_z \be_z ,
\eeq
where we have used cylindrical polar coordinates in $\bk$ space. This formula illustrates a key point. For the positive helicity states ($\alpha=+1$) which are of lower energy, the expectation value of velocity {\em vanishes} for $\bk = (\lambda/\sqrt{2}, 0, 0)$. This is the result of the fact that although the canonical momentum of the low energy states is of the order of $\lambda$, the {\em mechanical momentum} which is proportional to the velocity is small. It is the mechanical momentum that contributes to the stress tensor. The reason for the fall in pressure is now evident. In the presence of the gauge field, all the states with $+$ helicity have a smaller velocity (mechanical momentum). Since in the presence of the gauge field, there are always more occupied states with $+$ helicity, the pressure is expected to fall with increasing $\lambda$ for a given density $\rho$ of particles.

\begin{figure}
\centerline{\includegraphics[width=\myfigwidth]{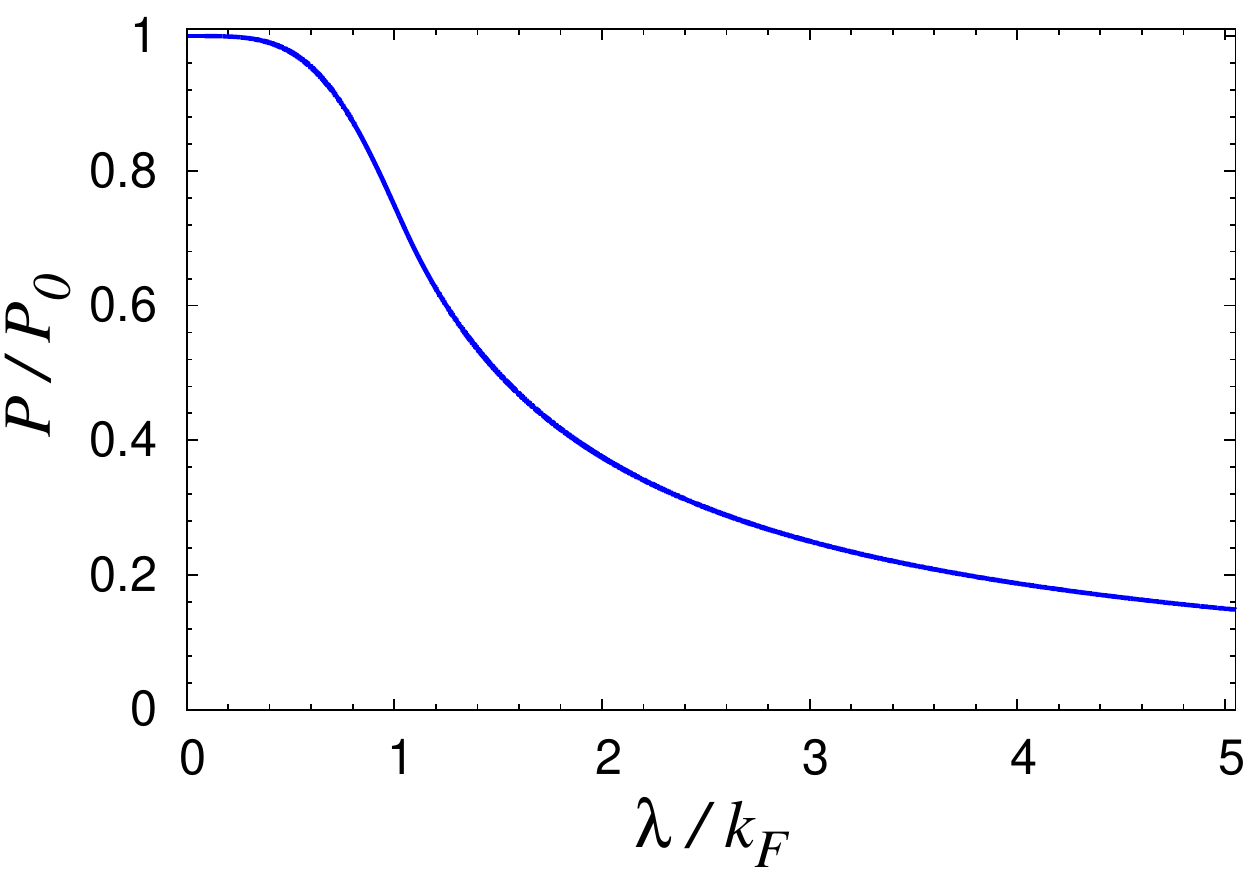}}
\caption{(Color online) Dependence of the pressure $P$ of a homogeneous gas of fermions in an EO gauge field on the strength of the gauge coupling $\lambda$. $P_0$ is the pressure of the free Fermi gas (no gauge field), and $\kf$ is the Fermi wave vector determined by the density.}
\mylabel{fig:EOPressure}
\end{figure}

\Fig{fig:EOPressure} illustrates the dependence of the pressure $P$ on the gauge coupling $\lambda$ for the EO gauge field calculated using \eqn{eqn:StressTensorExpect} and \eqn{eqn:Pressure}. The pressure falls as expected. In fact, for $\lambda > \lambda_T$ ($\lambda_T = \left( \frac{8 \sqrt{2}}{3 \pi} \right)^{1/3} \kf$, the gauge coupling at which the Fermi surface undergoes a topological transition, $\kf = (3 \pi^3 \rho)^{1/3}$ where $\rho$ is the density), we obtain 
\beq\label{eqn:EOPressure}
\frac{P}{P_0} = \frac{5 \sqrt{2}}{3 \pi} \frac{\kf}{\lambda} \;\;\;\; \mbox{for} \;\;\;\; \lambda \ge \lambda_T.
\eeq
We also note that the stress tensor calculated via \eqn{eqn:StressTensorExpect} is isotropic for this GFC.

Similar physics is at play also in the spherical gauge field where $\frac{P}{P_0} \sim \left(\frac{\kf}{\lambda}\right)^{4}$ for gauge couplings larger than that required to produce the change in topology of the Fermi surface. The more rapid fall of pressure with increasing gauge coupling owes to the fact that there are many more low energy states with vanishing velocity in the spherical gauge field.

These considerations now provide a clear physical picture of the shrinking of the cloud with increasing gauge coupling.

\subsection{Cloud Shape}

Within LDA the sole effect of the gauge field is the shrinking of
the cloud and cloud shape remains spherical for this isotropic trap.
For the EO gauge field this is a somewhat surprising result since one
would expect the one body wave function to have different behavior
``in-plane'' of the gauge field ($x-y$ plane) and ``out of plane''
($z$ direction).  To investigate this, we calculated the density
profile using the exact wave functions for this problem obtained
numerically using the method given in appendix
\ref{sec:NumericalCalculation} and compared this with the LDA result.

\begin{figure}
\centerline{\includegraphics[width=\myfigwidth]{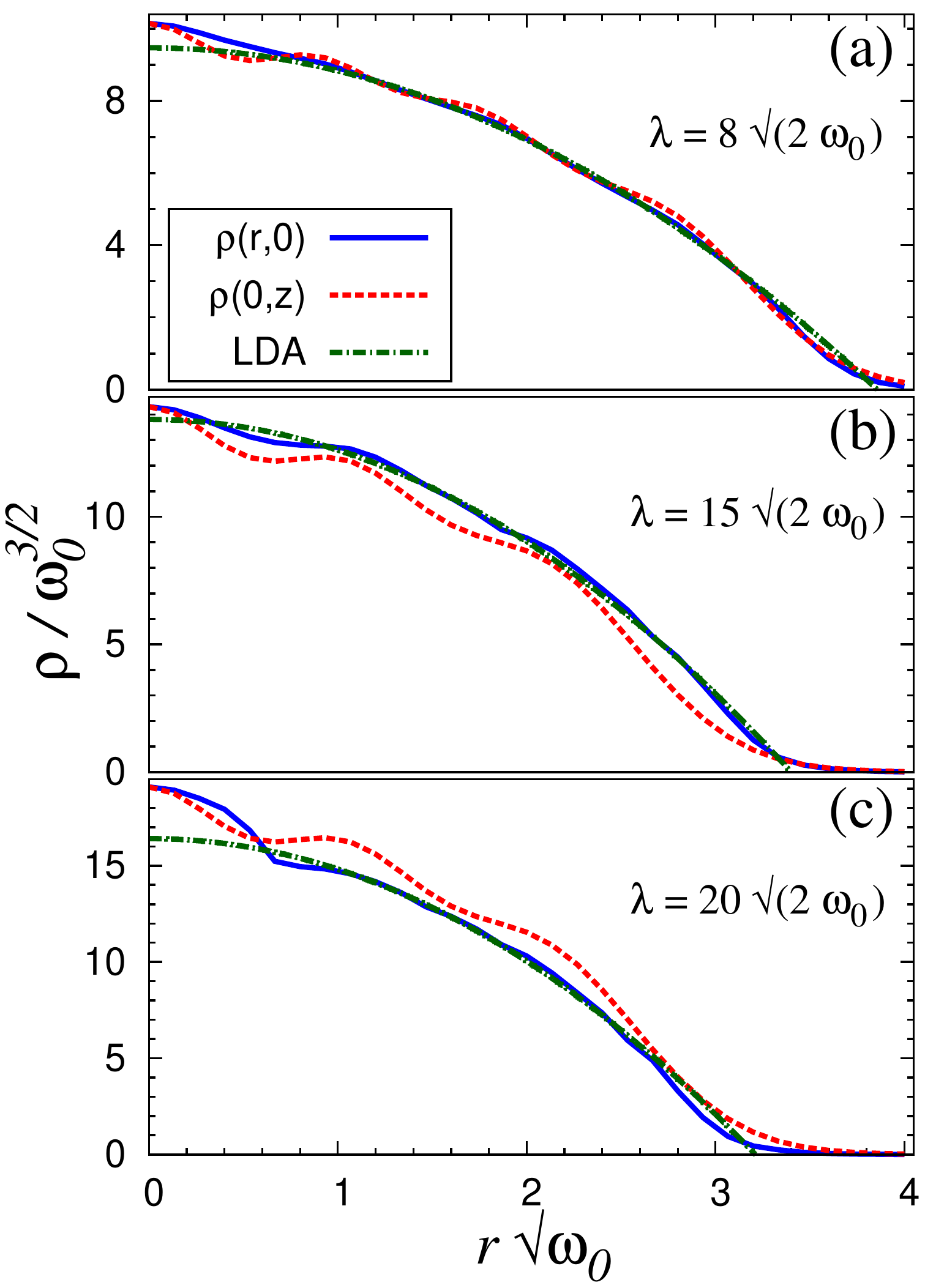}}
\caption{(Color online) Density profile of a gas of 910 trapped fermions in an EO gauge field. The in-plane density profile is indicated by $\rho(r,0)$ and out of plane profile by $\rho(0,z)$. }
\mylabel{fig:CloudShape}
\end{figure}

\Fig{fig:CloudShape} shows the density profiles obtained using the
exact numerical calculation of appendix \ref{sec:NumericalCalculation}
and that obtained from LDA for three values of gauge coupling $\lambda$
for the EO gauge field. For this gauge field, the symmetry of the
problem ensures that the density is a function only of the
in-plane radial coordinate $r$ and the $z$ coordinate, i.~e,
$\rho=\rho(r,z)$. We show curves corresponding to in-plane density
profile $\rho(r,0)$ and out of plane profile $\rho(0,z)$ obtained
from the exact numerical calculation. The results show a remarkable
feature. For $\lambda = 8 \sqrt{2 \omegazero}$ we see that in-plane and
out of plane density profiles are close to each other and agree quite
well with the LDA result. However, at a larger value of the gauge coupling  $\lambda = 15 \sqrt{2 \omegazero}, \; 20\sqrt{2\omegazero}$, we see that the in-plane density agrees
quite well with the LDA result. However, in both of these cases, the out of plane density {\em
systematically} deviates from the in-plane (and LDA) results. 

This effect can be understood by using the one particle spectrum given
in \eqn{eqn:EOEigval} obtained using our adiabatic theory. Filling up
of the levels associated with the quantum numbers $(n,m,n_z)$ occurs
by the ordering of their energies. When $\lambda$ is large, the levels
of the type $(0,m,0)$ are the lowest lying ones with the spacing
between them approximately equal to $2 m
\frac{\omegazero^2}{\lambda^2}$. On the other hand occupation of a
state with $n_z = 1$ or $n = 1$ requires an energy of order
$\omegazero$ higher than the ground state energy. Thus the number of
states with $n=n_z=0$ will be determined by the maximum value
of $m$ attained, $m^{max} \sim 2 \lambda/\sqrt{\omegazero}$. If $N_p$ is the
number of particles, we can estimate the maximum value of
$n_z$ of all occupied states. A simple calculation shows that
$n_z^{max} \approx \sqrt{\frac{\sqrt{\omegazero} N_p}{\lambda}}$. For
  the smallest value of $\lambda$ shown in \Fig{fig:CloudShape} the
  value of $m^{max}$, $n^{max}_z$ etc. are all approximately equal and
  the cloud is approximately isotropic with reasonable agreement with
  LDA. However, for the higher values of $\lambda$ in
  \Fig{fig:CloudShape}, $n^{max} \approx 6$ but $m^{max}$ is much
  larger. The reason why $\rho(r,0)$ agrees with LDA is now
  evident. Contributions to $\rho(r,0)$ arises from many states with
  different $m$ values and therefore the averaging/smoothing of the
  density is more prominent. This is not the case for $\rho(0,z)$
  where the density contribution is obtained only using $n_z^{max}$ number of
  $z$ wave functions. Not surprisingly, the out of plane density
  profile $\rho(0,z)$ shows large deviations from the LDA
  result. Finally, we address the reason of why $\rho(r,0)$ falls
  below $\rho(0,z)$ for larger $\lambda$. As $\lambda$ increases, the
  in-plane wave functions are ``more localized'' near the origin. This
  is because the low energy states are predominantly constructed out
  of plane waves with in-plane momentum of order $\lambda$. This means
  that real space wave function of the particle corresponds to
  that in which the particle is localized over a radial distance of order
  $\frac{1}{\lambda}$. These arguments suggest: a) the trap center
  density (for a given number of particles) scales with $\lambda$, b)
  the cloud will become more and more cigar shaped extending in the z direction as $\lambda$
  increases. Indeed, our results for the trap center density for larger $\lambda$s do agree with this argument, and for very high value of $\lambda$, we do
  find the tendency for the cloud to become cigar shaped. Quite
  interestingly, the density at a given distance
  from the trap center can be up to 10\% different along the in-plane and the
  out of plane directions. We believe that this systematic anisotropy
  should be observable and measurable in experiments.

\section{Summary and Conclusions}
\mylabel{sec:Summary}

In this paper we have investigated the physics of trapped
noninteracting fermions in the presence of a non-Abelian gauge field
that induces a generalized Rashba spin-orbit interaction. Specifically:
\begin{enumerate}
\item An adiabatic approximation is developed to obtain  the one particle levels in a trapping potential for a generic gauge field. The effective adiabatic Hamiltonian for the particle includes a gauge potential arising due to the induced connection from the internal spin degree of freedom. For the spherical gauge field this gauge potential is shown to be equivalent to the field of a monopole at the origin in the momentum space. Approximate analytic formulae for the levels are obtained for high symmetry gauge field configurations of interest. For the extreme oblate GFC these results are compared with the exact results obtained numerically and excellent agreement is demonstrated.
\item The effect of the gauge field on the size of the trapped cloud of fermions is investigated within a local density approximation. It is shown, quite generically, that the cloud shrinks with the increasing strength of the gauge field. Formulae for the dependence of the cloud size on the strength of the gauge field are obtained. The physical origins of this phenomenon are elucidated by our analysis of the stress tensor.
\item The density profile obtained using the local density approximation is compared with that obtained from a exact numerical calculation. It is shown that for the EO GFC, the density distribution is anisotropic in a characteristic way and should be observable in experiments.
\end{enumerate}

We believe that the results obtained in the paper will be useful to design experiments involving fermions in non-Abelian gauge field. In addition, our results also open doors to new possibilities with cold atoms that arises from using the non-Abelian gauge field in conjunction with a trapping potential or other types of potentials. In particular, we have shown  that a spherical non-Abelian gauge field and a harmonic trapping potential can be used to simulate the spherical geometry quantum hall system. Interestingly, the quantum hall system in this case is realized in the momentum space.

\appendix

\section{Numerical Calculation of One Particle States for the EO GFC}
\mylabel{sec:NumericalCalculation}

In this appendix we outline the method used to numerically obtain the
exact one particle states for trapped fermions in an EO
gauge field.

As is evident we need to focus only on the in-plane
degrees of freedom which are most conveniently described using polar
coordinates $(r,\phi)$. The Hamiltonain (\eqn{eqn:FullHamiltonian}) is rewritten in a more convenient form
\beq\mylabel{eqn:ExactHamiltonian}
{\cal H} = \underbrace{\frac{\omegazero^2}{2} \hat{\br}^2 + \frac{\hat{\bp}^2}{2}}_{{\cal H}_0} \underbrace{- \frac{\lambda}{\sqrt{2}} \left( \hat{p}^+ \tau^- + \hat{p}^- \tau^+ \right)}_{{\cal H}_\lambda}
\eeq
where $\hat{p}^\pm = \hat{p}_x \pm i \hat{p}_y$ and $\tau^{\pm} = \half (\tau_x \pm i \tau_y)$. In the following, energy will be measured in the units of $\omegazero$, distance in units of $1/\sqrt{\omegazero}$, momentum and gauge coupling in the units of $\sqrt{\omegazero}$, rendering all quantities dimensionless.

 The eigenstates of ${\cal H}_0$ are such that
\beq
{\cal H}_0 \ket{m,n,\sigma} = (2n +|m| + 1) \ket{m,n,\sigma}
\eeq
where $m$ is the angular momentum quantum number and $n$ is the radial quantum number, and $\sigma$ is the spin state which is $\uparrow$ or $\downarrow$, i.e., 
\beq
\ket{m,n,\sigma} = \ket{m,n}\otimes\ket{\sigma}
\eeq
with 
\beq
\braket{r,\phi}{m,n} = \sqrt{\frac{n!}{\pi (n + m)!}} e^{-\frac{r^2}{2}}r^m L^m_n(r^2) e^{i m \phi}
\eeq
where $L^m_n$ are associated Laguerre polynomials.

Since $J_z = L_z + \half \tau_z$ commutes with the Hamitlonian (\eqn{eqn:ExactHamiltonian}), we construct energy eigenstates which are also eigenstates of $J_z$. Noting that state $\ket{m,n,\uparrow}$ and $\ket{m+1,n,\downarrow}$ are eigenstates of $J_z$ with the eigenvalue $m+\half$, we can write the eigenstates of Hamiltonian \prn{eqn:ExactHamiltonian} as
\beq \label{eqn:StateExpansion}
\ket{\Phi} = \left( \begin{array}{c}
\sum_n a_{mn} \ket{m,n,\uparrow} \\
\sum_n b_{mn} \ket{m+1,n,\downarrow} 
\end{array}
 \right)
\eeq
where $a_{mn}$ and $b_{mn}$ are coefficients to be determined. This leads to a matrix eigenvalue problem which is completely defined with the specification of the following matrix elements:
\begin{widetext}
\beq
\begin{split}
\bra{m+1,n,\downarrow}{\cal H}_\lambda \ket{m,n',\uparrow} & = -i \frac{\lambda}{\sqrt{2}} \left( \sqrt{m+n+1} \, \delta_{n,n'} + \sqrt{n+1} \, \delta_{n,n'-1}\right) \\
\bra{m,n,\uparrow}{\cal H}_\lambda\ket{m+1,n',\downarrow} & = i \frac{\lambda}{\sqrt{2}} \left(
\sqrt{m+n'+1} \, \delta_{n,n'} + \sqrt{n'+1} \, \delta_{n',n-1} \right)
\end{split}
\eeq
\end{widetext}

In actual calculations the sum in \eqn{eqn:StateExpansion} has to be truncated at an upper value $N$ of $n$. We have chosen $N$ in such a way that the energy eigenvalues are converged to the desired double precision accuracy. When $\lambda = 20 \sqrt{2}$, we used $N=180$ to obtain this accuracy.

\bibliography{refRashbon_nagf}

\end{document}